\documentclass{emulateapj}

\usepackage{graphicx}
\usepackage{subfigure}
\usepackage{latexsym}

\newcommand{\HI}{\protect\ion{H}{1}}
\newcommand{\HII}{\protect\ion{H}{2}}
\newcommand{\msun}{$M_\odot$}
\newcommand{\lsun}{$L_\odot$}
\newcommand{\etal}{{et~al.}}
\newcommand{\mhi}{$M_{HI}$}
\newcommand{\kms}{km\,s$^{-1}$}
\newcommand{\mdyn}{$M_{dyn}$}
\newcommand{\lb}{$L_B$}

\begin{document}

\title{\HI\ Observations of SA 68-6597: the faintest Blue Compact Dwarf 
Galaxy.}
\author{D.J. Pisano\altaffilmark{1,2,3}, David C. Koo\altaffilmark{1,4,5}, 
Christopher N.A. Willmer\altaffilmark{1,4}, 
Kai Gerhard Noeske\altaffilmark{1,4},and
A.C. Phillips\altaffilmark{1,4}}
\altaffiltext{1}{Email: pisano@nrl.navy.mil, koo@ucolick.org, 
cnaw@ucolick.org, kai@ucolick.org, phillips@ucolick.org} 
\altaffiltext{2}{Naval Research Laboratory, Remote Sensing Division, 
Code 7213, 4555 Overlook Ave. SW, Washington, DC 20375}
\altaffiltext{3}{National Research Council Postdoctoral Fellow}
\altaffiltext{4}{UCO/Lick Observatory, University of California, Santa Cruz, 
CA 95064}
\altaffiltext{5}{Dept. of Astronomy and Astrophysics, University
of California, Santa Cruz, CA 95064}

\slugcomment{Accepted for Publication in ApJ Letters}
\accepted{2005 July 25}

\begin{abstract}

Blue compact dwarf galaxies (BCDs) are faint (M$_B\le$-17 mag) compact (R$<$1
kpc), at least qualitatively very blue galaxies due to active star
formation, and  have low metallicities.  Found serendipitously as part
of a redshift survey of faint galaxies with the Keck Telescope (DEEP), 
SA 68-6597
is at a distance of 80 Mpc, and is one of the faintest, -12.4 mag,
lowest metallicity, $\sim$0.05 Z$_\odot$, BCDs known.  Its H$\beta$ 
linewidth of $\sigma =$27 \kms\ and small size,
R$_{eff}\sim$190 pc, suggest that it is an extremely low mass galaxy.
We have used the Arecibo telescope to measure the \HI\ properties of
SA 68-6597 in order to better constrain its total mass and its
potential for future star formation.  SA 68-6597 has a
\mhi=(1.4$\pm$0.4)$\times$10$^7$\msun\ and an \HI\ FWHM linewidth of
33$\pm^{60}_{12}$.  Combining the \HI\ linewidth with an estimate of the 
size of the \HI\
disk, we derive a \mdyn$\gtrsim$3$\times$10$^7$\msun.  The
\mhi/\lb=1.0$\pm$0.3\msun/\lsun, \mdyn/\lb$\ge$2\msun/\lsun\ and 
\mhi/\mdyn$\lesssim$0.47 values are typical for BCDs.  
Combining the measured star formation rate of 0.003
\msun/yr with the \HI\ mass, we derive a gas depletion timescale of
5$\pm$2 Gyr.  While SA 68-6597 is a fainter, lower-mass, higher
metallicity counterpart to other BCDs like I Zw 18 and SBS 0335-052, 
its \HI\ properties suggest it will not evolve dramatically
in the near future.  Given the limits on its gaseous and dynamical
masses, SA 68-6597 may be able to evolve into a moderately massive 
dwarf spheroidal galaxy.
\end{abstract}


\section{Introduction}
\label{intro}

Blue Compact Dwarfs (BCDs) are faint \citep[M$_B\le$-17 mag,
e.g.][]{kon02},   compact \citep[diameters of the high surface
brightness regions of less  than 1 kpc, e.g.][]{thu81}, and blue
enough to suggest active star formation \citep[e.g.][]{gor81,thu81}.
They are typically low  metallicity systems \citep{izo99}.  Two of the
most extreme BCDs, in terms of luminosity, mass, and metallicity, are
I Zw 18 and SBS 0335-052.   I Zw 18 and SBS 0335-052 have luminosities
of -12.8 mag and -14.3 mag, total masses of 10$^{8.5-9.5}$\msun, HI
masses of 10$^{7.8-9.3}$\msun\ \citep*{vz98a,pus01}, and oxygen
abundances of  the ionized gas of $12+log(O/H) =$ 7.17 \& 7.34
\citep{izo99}--the lowest  known in the universe.  Because of these
properties, I Zw 18 and SBS 0335-052 are believed to be undergoing
early bursts of star formation \citep{izo04,lip99}.  While much
more luminous, BCD-like, {\sc{HII}}  galaxies at moderate redshift may
evolve into galaxies like NGC 205 \citep{koo94,koo95,guz96}, these
low luminosity, low mass BCDs may be the progenitors of dwarf
spheroidal galaxies like Carina.

SA 68-6597 was discovered serendipitously during the first
DEEP\footnote{Deep Extragalactic Evolutionary Probe:  see URL
\url{http://deep.ucolick.org}} run using the Keck LRIS
instrument\citep{oke95} with a 1200 l/mm grating (Koo et al. 2005, in
preparation).  This galaxy was selected because of its blue color and
visually estimated compact non-stellar appearance and faint apparent
magnitude. The LRIS data show that SA 68-6597 has a redshift of
$z$=0.0186 implying that it is intrinsically extremely faint.
Assuming a Hubble constant of 70 \kms\ Mpc$^{-1}$,  SA 68-6597 is
located at a distance of 80 Mpc and has a B magnitude of -12.4 mag as
measured by the DEEP.team.   Using the combination of the
\ion{O}{3}$[\lambda 5007]$/H$\beta$ and the  \ion{N}{2}$[\lambda
6583]$/H$\alpha$ line ratios as measured with HIRES,  Koo et al. find
that SA 68-6597 has an extremely low excitation temperature,
$\sim$14,500 K, and metallicity $12+log(O/H) \sim$7.4 ($\sim$0.05
Z$_\odot$).  This places SA 68-6597 well away from the well-defined
locus of \HII\  galaxies \citep[e.g.][]{lee04} and Local Group dwarf
irregulars \citep{mat98}  in this parameter space.   The most similar
galaxy to SA 68-6597 in this space is the BCD SBS 0335-052.
Follow-up observations (Koo et al. 2005, in preparation) with Keck
HIRES \citep{vog94} also show that the emission lines have Gaussian
velocity dispersions of $\sim$ 27 km/s.    HST WFPC2 images reveal a
very small galaxy, R$_{25}$ of 
1.0$\pm^{0.1}_{0.05}$\arcsec = 400$\pm^{40}_{20}$ pc.  Combined with
the small optical linewidth, this implies that SA 68-6597 is a  very
low mass galaxy, $\sim$10$^7$\msun.  Based on the flux of the
H$\alpha$ line in the LRIS spectra, the star formation rate of this galaxy, 
0.003 \msun\ yr$^{-1}$, is similar to what is expected for a BCD given its
inferred low total mass \citep{hop02}.  These properties from Koo et al. 
(2005, in preparation) and summarized in Table~\ref{props}, strongly suggest 
that SA 68-6597 is a blue compact dwarf that is fainter and smaller than 
other BCDs \citep[e.g.][]{thu81,sal02}.   It is a fainter, lower-mass, but 
slightly higher metallicity counterpart to the more famous BCDs I Zw 18 and
SBS 0335-052.  It is the faintest known BCD (Koo et al. 2005, in
preparation).  Its extreme nature makes this galaxy a  particularly
interesting probe of low mass galaxy formation  \citep[see][and
references therein for a discussion]{pus01}.

Observations of the 21-cm line of neutral hydrogen (\HI) can help us
better understand the nature of SA 68-6597.   The \HI\ content of SA
68-6597 is an important measure of its potential for future star
formation.  The current burst of star formation is small in an
absolute sense, 0.003 \msun\ yr$^{-1}$, but if the \HI\ mass  is also
low then it can still rapidly consume its \HI\ and subsequently
passively evolve, fade and possibly become a galaxy like the Carina
dwarf spheroidal in the Local Group.  If the \HI\ mass  is much
higher, then it is more likely to continue forming stars for a long
time and will retain its current appearance.  Furthermore, while the
optical emission lines have a width of 27 \kms, this linewidth may not
trace the entire gravitational potential of the galaxy.  The \HI\ gas
tends to trace the gravitational potential to larger radii than the
stars or ionized gas.  In addition, ionized gas may be tracing
galactic outflows; this is less likely for the neutral gas.   For all
of these reasons, \HI\ provides the best measure of the total,
dynamical mass of a galaxy.  The dynamical mass is an important
constraint on the evolution of a galaxy.  If SA 68-6597 has a low
total mass, then it may eject its neutral gas before it can consume it
in star formation \citep[e.g.][]{mac99}.  If it is higher,  then it
may be too massive to evolve into a dwarf spheroidal galaxy.  In this
paper, we report on our Arecibo observations of  \HI\ in SA 68-6597.
These observations help constrain the potential for future star
formation in the galaxy and provide a more robust measure of the total
mass of the galaxy constraining the current nature and future
evolution of SA 68-6597, and its relation to other BCDs like I Zw 18.

\section{Arecibo \HI\ Observations \& Reductions}
\label{obs}

We observed SA 68-6597 with the Arecibo\footnote{The Arecibo
Observatory  is part of the National Astronomy and Ionosphere Center
which is operated  by Cornell University under a Cooperative Agreement
with the National  Science Foundation.} 305 m telescope on 2004 August
1--4 and October 3--4.   We observed only at night to minimize solar
interference.  We used the L-wide  receiver for all our observations.
This receiver has a system  temperature of $\sim$27 K, and a gain of
$\sim$10 K Jy$^{-1}$ as measured by the Arecibo staff.  Both values
are weakly dependent on the zenith angle of the observation.  Data
were processed through the interim  correlator in both linear
polarizations over total bandwidths of 25 MHz and 12.5 MHz,
corresponding to a velocity range of $\sim$5000  \kms\ and
$\sim$2500~\kms.  Each bandwidth and polarization had 9-level sampling
and 2048 channels, resulting in a velocity resolution per channel of
5.2 \kms\ and 2.6 \kms.  Our observations utilized a high pass filter
to block interference below 1370 MHz contaminating our band.  The beam
size of the L-wide receiver according to the Arecibo documentation is
3.1$\arcmin \times$ 3.5$\arcmin$.  At the distance of SA 68-6597, 80
Mpc, this corresponds to a linear size of 72 kpc$\times$ 81 kpc.  As
the effective radius of the stellar emission is only 190 pc, this
should be  more than sufficient to encompass all of the \HI\
associated with this  galaxy.  Yet this beam size is small enough that
there are no known galaxies at a similar redshift that can
contaminate our \HI\ measurements; the closest galaxy is $\gtrsim$400
kpc away and there are no known groups within 3 Mpc based on a 
NED\footnote{The NASA/IPAC Extragalactic Database (NED) is operated by 
the Jet Propulsion Laboratory, California Institute of Technology, under 
contract with the National Aeronautics and Space Administration.} search.  
It is further unlikely that there
are any \HI-rich, optically invisible galaxies exist that could contaminate
our measurements \citep{doy05}.  

We used the standard position switching algorithm for our observations
spending 5 minutes on SA 68-6597 followed by 5 minutes offset to blank
sky by 5 minutes in right ascension such that we tracked the same
azimuth and zenith angle as the on-source scan.  This was repeated
for all six nights for an on-source integration time of 240 minutes.
Our data were reduced using standard Arecibo IDL routines written by
Phil Perrilat.  Each bandwidth of each scan was calibrated separately
and the polarizations were averaged together before a first or second
order baseline was fit across the portion of the spectrum clean of any
interference.  All scans were then averaged together to produce the
final spectrum.  While some scans showed signatures of interference
around 1400 MHz, since SA 68-6597 is at a frequency of 1394 MHz this
should not affect our ability to detect the galaxy.  The bandpass,
however, was significantly better for the 12.5 MHz band as a result of
the RFI at 1400 MHz, and so we proceeded only with this band.  Because
the two bands are split after the first  amplification, their noise is
not independent and, therefore, it would  not have helped to combine
these bands.  The resulting noise was 0.22 mJy per 2.6 \kms\
channel.  For analysis, we binned this spectrum by 4 channels to a
resolution of 10.8 \kms, improving the noise to 0.10  mJy per channel.
The observational details are listed in Table~\ref{props}.  Our flux
measurements of a bright galaxy, UGC 199, are within 17\% of
previously published values \citep{sch90}.  This is a much smaller
source of error than that of random noise and can be disregarded for
this work.

\begin{deluxetable}{@{}lrl}
\tablecolumns{4}
\tablewidth{0pc}
\tablecaption{Properties of SA 68-6597\label{props}}
\tablehead{\colhead{Parameter} & \colhead{Value} & \colhead{Units}}
\startdata
\hspace*{0em}Optical Properties\tablenotemark{a}:       & & \\
\hspace*{1em}Right Ascension (J2000)& 00:17:15.41 & h:m:s \\
\hspace*{1em}Declination (J2000)    & 15:48:38.36 & $^\circ : \arcmin\ : \arcsec\ $ \\
\hspace*{1em}Optical Heliocentric Velocity  & 5595$\pm$40 & \kms \\
\hspace*{1em}Distance                  & 80     &  Mpc \\
\hspace*{1em}Optical Velocity Dispersion & 27$\pm^1_2$ &  \kms \\
\hspace*{1em}Absolute B Magnitude     & -12.4 & mag \\
\hspace*{1em}Luminosity		   & 1.4  & 10$^7$\lsun \\
\hspace*{1em}R$_{25}$              & 400$\pm^{40}_{20}$ & pc \\
\hspace*{1em}Star Formation Rate    & 0.003     & \msun/yr \\
\hline
\hspace*{0em}Observational Properties\tablenotemark{b}: & & \\
\hspace*{1em}Time on Source         & 240  & min. \\
\hspace*{1em}Channel Width (binned) & 10.8 & \kms \\
\hspace*{1em}RMS noise per channel  & 0.10 & mJy  \\
\hline
\hspace*{0em}\HI\ Properties\tablenotemark{b}:          & & \\
\hspace*{1em}Peak Flux		   & 0.31 & mJy  \\
\hspace*{1em}Heliocentric Recession Velocity & 5557$\pm$5 & \kms \\
\hspace*{1em}Velocity Width (20\%)  & 51$\pm^{93}_{19}$ & \kms \\
\hspace*{1em}Integrated Flux        & 9.5$\pm$2.5 & mJy \kms \\
\hspace*{1em}\HI\ Mass		   & 1.4$\pm$0.4 & 10$^7$M$_\odot$ \\
\hspace*{1em}\mhi/\lb              & 1.0$\pm$0.3 & \msun/\lsun \\
\hspace*{1em}Gas Depletion Time	   & 5$\pm$2   & Gyr \\
\hspace*{1em}\mdyn                 & $\ge$3.0  & 10$^7$\msun \\
\hspace*{1em}\mdyn/\lb             & $\ge$2    & \msun/\lsun \\
\hspace*{1em}\mhi/\mdyn            & $\lesssim$0.47 &  \\
\enddata
\tablenotetext{a}{From Koo \etal\ (2005), in preparation}
\tablenotetext{b}{this paper}
\end{deluxetable}

\section{Results}
\label{results}

\begin{figure}[hb]
\plotone{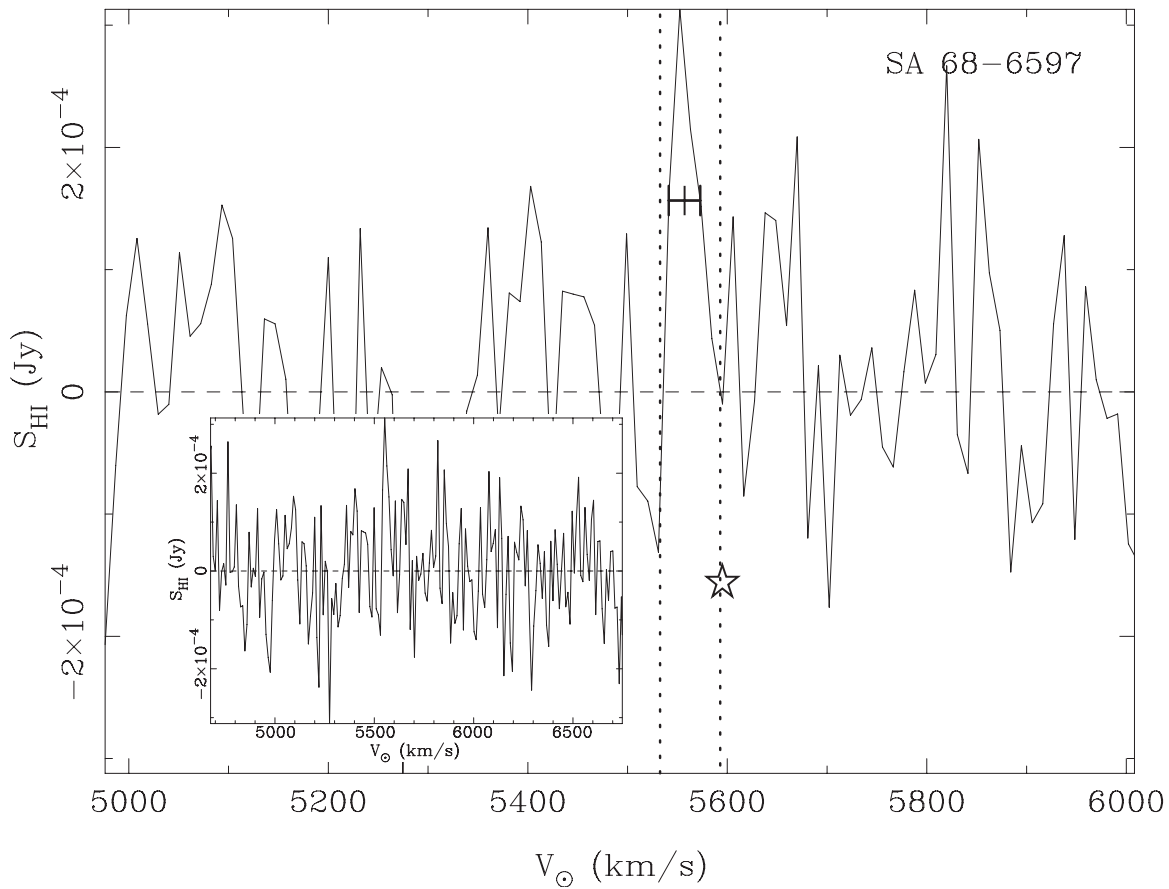}
\caption{A plot of a portion of the Arecibo \HI\ spectrum.  The dashed line 
indicates the origin, while the vertical dotted lines indicate the range over 
which the line was measured.   The vertical tick marks the \HI\ recession 
velocity of the galaxy, while the star indicates the optical recession 
velocity.  The horizontal solid line indicates the measured \HI\ FWHM.  
The inset shows the entire Arecibo bandpass (excluding the edges) illustrating 
that this line is the brightest feature in the spectrum.
\label{hispec}}
\end{figure}

Figure~\ref{hispec} shows the binned \HI\ spectrum of SA 68-6597 near 
the known optical velocity of the galaxy.  The small inset in the 
lower left corner shows the entire spectrum (excluding the edges of 
the bandpass).  The \HI\ emission is very weak, but clearly detected.  
The line has a peak flux of 0.31 mJy located at a velocity of 5552 \kms.  
This is only a 3$\sigma$ detection, but it is the brightest feature in 
the spectrum and is located within $\sim$40 \kms\ of the optical velocity
of SA  68-6597; which is within the 1$\sigma$ uncertainties of the
optical redshift.  To check its reality, we split the raw data into
various subsets (e.g. by sets of days, polarization, etc.)  and
searched for the line in these data.  The line was either visible in
the subsets or its absence was consistent with the noise levels; thus
we believe that it is a real emission line from SA 68-6597.

The vertical dotted lines in Figure~\ref{hispec} indicate the region
over which we measured the \HI\ properties of SA 68-6597.  The
integrated flux is 0.0095$\pm$0.0025 Jy \kms, which translates to a
\mhi\ of  $(1.4\pm0.4)\times$10$^7$\msun--a 3.8$\sigma$~ detection
over five channels.  We know of no galaxies within the 3.5$\arcmin$
beam of Arecibo that may be contaminating this measurement, so we
believe that this \HI\ is truly associated with SA 68-6597.  Combining
these \HI\ data with the optical properties, we find a  \mhi-to-\lb\
ratio of 1.0$\pm$0.3\msun/\lsun.  The gas depletion timescale,
$\tau$=\mhi/SFR, is 5$\pm$2 Gyr without accounting for helium,
molecular gas, or recycling.

The \HI\ linewidth at 50\% of the peak flux, W$_{50}$ (FWHM), is
measured to be 32$\pm$10 \kms, centered at 5557$\pm$5 \kms.  Again,
this is within the uncertainties of the optical recession velocity.
Because of the low  signal-to-noise ratio of the detection, this value
is highly imprecise and  potentially inaccurate.  To address this
issue, we have used a Monte Carlo  simulation of a Gaussian line with
a peak signal-to-noise ratio of  3$\sigma$ to relate the measured FWHM
to the true FWHM.  We find that the the true \HI\ FWHM =
33$\pm^{60}_{12}$ \kms.  Converting this to a W$_{20}$ yields
51$\pm^{93}_{19}$ \kms, assuming a Gaussian lineshape.

One of the main goals of our project is to determine the dynamical
mass,  \mdyn, of SA 68-6597 using the \HI\ line.  As discussed in
Section~\ref{intro}, the \HI\ line is generally believed to be a more
reliable tracer of the gravitational potential than the H$\beta$ line.
Because of the large  uncertainties associated with our \HI\
measurement, and the additional uncertainties from the unknown
inclination of SA 68-6597, we are practically restricted to calculating
a lower limit to \mdyn.  We follow the same procedure to derive the
dynamical mass as in \citet{pis01} using the following standard
formula assuming the \HI\ is in circular rotation: 
\mdyn$(<R)~\ge~V_{rot}^2 \times R/G$.


In this case we take V$_{rot}$ to be half of the lower limit on
W$_{20}$ uncorrected for inclination, for the radius we scale R$_{25}$ 
using a canonical scaling factor from \citet{bro97} to 
get R$_{HI}$=680$\pm^{210}_{200}$ pc.  If we use these values to calculate a 
lower limit to \mdyn, we find it is greater than 3.0$\times$10$^7$\msun.   
This yields \mhi/ M$_{dyn} \le$ 0.47, and \mdyn/\lb$\ge$2.   See \citet{pis01}
for a discussion of the uncertainties involved in this calculation.  All of  
these measured and derived properties of SA 68-6597 are summarized in
Table~\ref{props}.

\section{Discussion}
\label{discussion}

Our \HI\ observations have revealed that SA 68-6597 is not only a low
luminosity galaxy, but also has a low \mhi, and probably a low \mdyn\
as well.  By all three measures, SA 68-6597 reveals itself to be an
extreme cousin of other BCDs.  BCDs typically have
\mhi$\sim$10$^{8-9}$\msun, with only a few as low as 10$^7$\msun\ or as
high as 10$^{10}$\msun.   They have \mdyn$\sim$10$^{8-10}$\msun, and
\lb$\sim$10$^{8-10}$\lsun\  \citep{cha77,thu81,hof89,sta92,thu99,sal02}.  SA
68-6597 represents the extreme low mass, low luminosity
end of BCDs and is not a distinctly different class of galaxy as
generally evidenced by its scale-free properties, such as the
\HI-mass-to-light ratio, \mhi/\lb, and the gas-mass  fraction,
\mhi/\mdyn.

SA 68-6597 has an \mhi/\lb\ ratio that is consistent with that of the
large samples of BCDs studied by \citet{sta92,vz98b,vz01,sal02,hof03,thu04} 
who found ratios  ranging from $\sim$0.33-1.46 \msun/\lsun.  SA 68-6597's
\HI\ mass-to-light ratio is even within the  range for luminous
compact blue galaxies studied by \citet{gar04b}, but is  about twice
the median value.  Only the study of \citet{hof89} found a
significantly lower \mhi/\lb\ value for 11 Virgo cluster BCDs of
0.04 \msun/\lsun.  The \mhi/\mdyn\ values for BCDs are also
generally consistent with SA 68-6597's upper limit of 0.47.  
A variety of studies of BCDS find \mhi/\mdyn\ ratios ranging 
from 0.01-0.78 \citep{hof89,vz98b,hof03,thu04}.
The ratio of dynamical mass-to-light, \mdyn/\lb, for other BCDs ranges
from 0.18-2.62 \msun/\lsun\ \citep{hof89,sta92,thu04}, which is also
consistent with the lower limit of 2 \msun/\lsun\ for SA 68-6597.  It's
SFR and \mhi\ are consistent with expectations for BCDs based on SA 68-6597's
\HI\ linewidth \citep{hop02}.  Even 
luminous compact blue galaxies have a median \mdyn/\lb\ only slightly higher
(5 \msun/\lsun) than that of SA 68-6597\citep{gar04b}.   All of these ratios
suggest that SA 68-6597 is an extremely faint, extremely low-mass version of
a typical blue compact dwarf.

In terms of individual BCDs, SA 68-6597 is quite similar in its
gaseous  properties to I Zw 18 and Haro 4.  I Zw 18 is still slightly
more massive and  more luminous with a \mhi\ = 2.6$\times$10$^7$\msun,
\mdyn=2.6$\times$10$^8$\msun, and \lb\ = 3.5$\times$10$^7$\lsun\
\citep{vz98a}.  Nevertheless, with \mhi/\lb\ = 0.7 and  \mdyn/\lb\ = 5
I Zw 18 has mass-to-light ratios nearly identical to SA  68-6597.  Its
gas-mass fraction of 0.1 is also similar to SA 68-6597.   Haro 4
is slightly less similar with \mhi\ = 2$\times$10$^7$\msun, \mhi/\lb\
= 0.17\msun/\lsun, \mdyn=5$\times$10$^8$\msun, \mdyn/\lb\ = 4.8\msun/\lsun, 
and \mhi/\mdyn =0.03 \citep{bra04}.  While \mhi, \mdyn, and \mdyn/\lb\ of 
Haro 4 are close to those of SA 68-6597, the \mhi/\lb\ is lower, and
the gas-mass fraction is lower than SA 68-6597, but still consistent
with it.

The question is then raised: ``what do these properties say about the
evolutionary path of SA 68-6597?''  Our derived gas depletion
timescale,  $\tau$, for SA 68-6597 is 5$\pm$2 Gyr.  This value
provides a rough measure  of the time it will take for SA 68-6597 to
consume all of its gas at its current rate of star formation
\citep{ken83}.  This is similar to many of the measured values for a 
sample of 21 BCDs from \citet{hop02}, but is much greater than a 
sample of 15 BCDs studied by \citet{sag92}.  It has a gas depletion
timescale equal to the median value for the sample of field and cluster 
galaxies studied by \citet{ken83}. SA 68-6597 has a larger
$\tau$ than either I Zw 18 or SBS 0355-052 \citep{hop02}.
It has a shorter $\tau$ than all Local Group dwarf irregular galaxies
except IC 10 and NGC 6822 \citep{mat98}.  Our estimate does not account 
for the contribution of helium, recycling 
of gas, molecular gas, a decreasing SFR or less than 100\% star formation
efficiency--all of which  would increase $\tau$--or the possible
effects of outflow--which would  decrease $\tau$.  \citet{mac99} suggest
that galaxies with gas masses below 10$^6$\msun\ can suffer complete
blowout of their gas, while galaxies between 10$^6$ and 10$^7$\msun\ may
suffer partial blowout.  Because the derived gas mass of SA 68-6597 is 
$\sim$10$^7$\msun, we expect that it could only have a partial outflow 
\citep{mac99}.  The results of \citet{mac99} are based on a dark matter
halo approximately 10-100$\times~$ larger than the gas mass.  
Overall, this means that, SA 68-6597
should evolve in a similar fashion to many less extreme BCDs and 
normal spiral and irregular galaxies; it will not rapidly consume
its gas and passively evolve in the near future.  The large \mdyn\
of SA 68-6597 implies that if and when it consumes all its gas, it
may be able to evolve into a moderate mass analog of the Local Group 
dwarf spheroidals \citep{mat98}.

While the low signal-to-noise ratio of our observations make our
velocity  widths very uncertain, it is worth noting a potentially
interesting property  of SA 68-6597.  The H$\beta$ velocity dispersion
of SA 68-6597 is  27$\pm^1_2$ \kms, while the \HI\ dispersion is
14$\pm^{25}_{5}$ \kms.  If the \HI\ linewidth is actually smaller than
that of the ionized gas, then we may be seeing evidence of a galactic
outflow in SA 68-6597.  Such outflows could result in partial blowout
of the ISM in the galaxy of anywhere between $\lesssim$1\% to 100\% 
\citep{mac99} potentially reducing the gas depletion time.  Outflows 
can also be particularly efficient at
ejecting metals into the intergalactic medium \citep{mac99,fer00}
permitting a galaxy like SA 68-6597 to form many generations of stars
while maintaining its very low metallicity.  Clearly more sensitive
and detailed \HI\ observations are needed to address this issue.

\section{Conclusions}
\label{conc}

We have observed the recently discovered blue compact dwarf galaxy, SA
68-6597, with the Arecibo telescope to determine its gas content and
better constrain  its total mass using the \HI\ line.  SA 68-6597 is
one of the faintest, lowest metallicity BCDs known.  

SA 68-6597 has properties which indicate it is a typical blue compact
dwarf galaxy in all ways, except for its extremely low luminosity and
small \HI\ and dynamical masses.  In this way it represents the faint,
low mass tail of the distribution of BCD properties.  It is slightly
fainter and less massive than the famous BCD I Zw 18 and only
slightly more metal rich.  SA 68-6597's gas depletion timescale is similar
to the value for other BCDs and normal field galaxies, yet is shorter than 
most dwarf irregulars in the Local Group.  Nevertheless, SA 68-6597 can 
continue to form stars at  its current, prolific rate for almost 5 Gyr and 
would, therefore, be unlikely to fade significantly in that time.  When
it does fade, its relatively large dynamical mass suggests it may be able
to evolve into a massive dwarf spheroidal galaxy.  Its future evolutionary 
path remains murky.  

Because of the combination of the distance and low \HI\ mass of SA
68-6597,  our detection was only at the 3$\sigma$ level, meaning that
the measured  linewidth and derived dynamical mass are poorly
constrained.  Nevertheless,  the most probable value of the \HI\
linewidth is less than the measured  H$\beta$ linewidth indicating
that galactic outflows may be present in SA 68-6597.  Because of the
potential implications of such a situation on SA 68-6597's evolution,
more  sensitive, spatially resolved \HI\ observations of SA 68-6597
are essential to unravel its current nature and reveal its evolutionary path.

\acknowledgements

This research has made use of the NASA/IPAC Extragalactic Database
(NED) which is operated by the Jet Propulsion Laboratory, California
Institute of Technology, under contract with the National Aeronautics
and Space Administration.  The authors wish to thank the staff at
Arecibo, especially Tapasi Ghosh, for their help with our
observations.  We also wish to thank the Arecibo  observatory for
granting us more observing time to improve the quality of our
observations.  We thank Eric Wilcots for assistance with the observing
and helpful discussions.  This research was performed while
D.J.P. held a  National Research Council Research Associateship Award
at the Naval Research  Laboratory.  Basic research in astronomy at the
Naval Research  Laboratory is funded by the Office of Naval Research.
D.J.P. also acknowledges generous support from the ATNF via a Bolton
Fellowship and from NSF MPS Distinguished International Research
Fellowship grant AST 0104439.  D.C.K. acknowledges support from NSF
AST-0071198  and HST GO-07339.01-96A.



\end{document}